

\magnification 1200

\font\eightrm=cmr8
\font\eighti=cmmi8
\font\eightsy=cmsy8
\font\eightbf=cmbx8
\font\eighttt=cmtt8
\font\eightit=cmti8
\font\eightsl=cmsl8
\font\sixrm=cmr6
\font\sixi=cmmi6
\font\sixsy=cmsy6
\font\sixbf=cmbx6
\catcode`@11
\newskip\ttglue
\font\grrm=cmbx10 scaled 1200

\def\eightpoint{\def\rm{\fam0\eightrm}
\textfont0=\eightrm \scriptfont0=\sixrm \scriptscriptfont0=\fiverm
\textfont1=\eighti \scriptfont1=\sixi \scriptscriptfont1=\fivei
\textfont2=\eightsy \scriptfont2=\sixsy \scriptscriptfont2=\fivesy
\textfont3=\tenex \scriptfont3=\tenex \scriptscriptfont3=\tenex
\textfont\itfam=\eightit \def\it{\fam\itfam\eightit}
\textfont\slfam=\eightsl \def\sl{\fam\slfam\eightsl}
\textfont\ttfam=\eighttt \def\tt{\fam\ttfam\eighttt}
\textfont\bffam=\eightbf
\scriptfont\bffam=\sixbf
\scriptscriptfont\bffam=\fivebf \def\bf{\fam\bffam\eightbf}
\tt \ttglue=.5em plus.25em minus.15em
\normalbaselineskip=6pt
\setbox\strutbox=\hbox{\vrule height7pt width0pt depth2pt}
\let\sc=\sixrm \let\big=\eightbig \normalbaselines\rm}
\newinsert\footins
\def\newfoot#1{\let\@sf\empty
  \ifhmode\edef\@sf{\spacefactor\the\spacefactor}\fi
  #1\@sf\vfootnote{#1}}
\def\vfootnote#1{\insert\footins\bgroup\eightpoint
  \interlinepenalty\interfootnotelinepenalty
  \splittopskip\ht\strutbox 
  \splitmaxdepth\dp\strutbox \floatingpenalty\@MM
  \leftskip\z@skip \rightskip\z@skip
  \textindent{#1}\footstrut\futurelet\next\fo@t}
\def\fo@t{\ifcat\bgroup\noexpand\next \let\next\f@@t
  \else\let\next\f@t\fi \next}
\def\f@@t{\bgroup\aftergroup\@foot\let\next}
\def\f@t#1{#1\@foot}
\def\@foot{\strut\egroup}
\def\footstrut{\vbox to\splittopskip{}}
\skip\footins=\bigskipamount 
\count\footins=1000 
\dimen\footins=8in 

\def\ref#1{$^{{#1})}$}
\def\flex{\raise 6pt\hbox{$\leftrightarrow $}\! \! \! \! \! \! }

\newbox\bigstrutbox
\setbox\bigstrutbox=\hbox{\vrule height10pt depth5pt width0pt}
\def\bigstrut{\relax\ifmmode\copy\bigstrutbox\else\unhcopy\bigstrutbox\fi}
\def\refer[#1/#2]{ \item{#1} {{#2}} }
\def\rev<#1/#2/#3/#4>{{\it #1\/} {\bf#2}, {#3}({#4})}
\def\boxit#1{\vbox{\hrule\hbox{\vrule\kern3pt
\vbox{\kern3pt#1\kern3pt}\kern3pt\vrule}\hrule}}

\def\2figure#1#2#3#4{\vbox{ \hrule width#1truecm \hbox{\vrule height#2truecm
\hskip #1truecm
\vrule height#2truecm }\hrule width#1truecm \hbox{\vrule\vbox{\hsize #1truecm
\baselineskip=10pt
\noindent\strut#3}\vrule}\hrule width#1truecm
\hbox{\vrule\vbox{\hsize #1truecm
\baselineskip=10pt
\noindent\strut#4}\vrule}\hrule width#1truecm  }}
\def\3figure#1#2#3#4#5{\vbox{ \hrule width#1truecm \hbox{\vrule height#2truecm
\hskip #1truecm
\vrule height#2truecm }\hrule width#1truecm \hbox{\vrule\vbox{\hsize #1truecm
\baselineskip=10pt
\noindent\strut#3}\vrule}\hrule width#1truecm
 \hbox{\vrule\vbox{\hsize #1truecm
\baselineskip=10pt
\noindent\strut#4}\vrule}
\hrule width#1truecm \hbox{\vrule\vbox{\hsize #1truecm
\baselineskip=10pt
\noindent\strut#5}\vrule}\hrule width#1truecm  }}

\def\sqr#1#2{{\vcenter{\hrule height.#2pt
   \hbox{\vrule width.#2pt height#1pt \kern#1pt
    \vrule width.#2pt}
    \hrule height.#2pt}}}


\def\ok{\overline k }
\def\ow{\overline w }
\def\oz{\overline z }
\def\ox{\overline X }
\def\atres{{\cal A}_3 }
\def\aquatro{{\cal A}_4 }
\def\smin{\,\raise 0.06em \hbox{${\scriptstyle \in}$}\,}
\def\smsubset{\,\raise 0.06em \hbox{${\scriptstyle \subset}$}\,}

\def\Natural{\hbox{\hskip 1.5pt\hbox to 0pt{\hskip -2pt I\hss}N}}

\def\Rational{\hbox{\hbox to 0pt{\hskip 2.7pt \vrule height 6.5pt
                                  depth -0.2pt width 0.8pt \hss}Q}}
\def\Real{\hbox{\hskip 1.5pt\hbox to 0pt{\hskip -2pt I\hss}R}}
\def\Complex{\hbox{\hbox to 0pt{\hskip 2.7pt \vrule height 6.5pt
                                  depth -0.2pt width 0.8pt \hss}C}}

\baselineskip=14truept plus 2truept
\lineskip=2truept plus 1pt minus1truept
\lineskiplimit=1truept
\parskip=1truept plus 1truept
\hsize 6truein
\vsize 8.5truein
\centerline{\grrm Tree Amplitudes in Noncritical N=2
Strings\newfoot{${}^\#$}{Talk
given in the Workshop on Superstrings and Related Topics (ICTP-Trieste)}}
\vskip 1cm
\centerline {D.Dalmazi\newfoot{*}{Supported by CNPq. }
\newfoot{${}^\dagger $}{E-mail: dalmazi@uspif.if.usp.br}}
\vskip .2cm
\centerline {Instituto de F\'\i sica, Univ. S\~ao Paulo, CP 20516,
S\~ao Paulo, Brazil}
\vskip 2cm
\centerline {\bf Abstract }
\vskip .2cm

Recent results for tree amplitudes for the $N=2$ noncritical strings
are presented and compared with the critical case. Arguments are
given which indicate a certain discontinuity in passing from
the $\hat c < 1$ model (in a Coulomb gas representation) to the
$\hat c = 1$ critical case.

\vfill\eject
\centerline{\grrm Tree Amplitudes in Noncritical N=2 Strings.}
\vskip .8cm
\centerline {D.Dalmazi}
\vskip .15cm
\centerline {Instituto de F\'\i sica, Univ. S\~ao Paulo, CP 20516,
S\~ao Paulo, Brazil}
\vskip .8cm

\centerline {\bf 1. Introduction }
\vskip .7cm

In the last few years noncritical strings have been extensively
studied from many different viewpoints. Specially exciting is the
nonperturbative aspect of the results furnished by the discrete
approach based on matrix models. The continuous approach
on the other hand has
proven to be less powerful for higher genus calculations
but the tree level (genus zero)
correlators\ref{1} of physical operators agree with the matrix model
results for the sphere for  $c=1$ (see [2] for a comparison).
For $c<1$ the spectrum of physical operators is not exactly the
same (see comments in [3]) but the scaling behaviour of the
correlators agree\ref{4} with
the results predicted by matrix models.
Although less powerful, the continuous approach is most easily generalized
to the supersymmetric strings. We have studied such
generalization in [5] for the case of the $N=1$ noncritical string by
calculating correlation
functions in a ${\hat c}\le 1 (c_m = {3{\hat c}\over 2})$
$N=1$ matter in a Coulomb gas representation conformally
coupled to a $N=1$ superliouville theory (see also [1,6]). The
results obtained in the NS-sector are very simple and similar
to the bosonic case ($N=0$). In the $N=0$ and
$N=1$ (NS-sector) cases the only propagating
particle is a massless scalar. Such particle is a remnant of the
tachyon ground state of the respective critical dimensions $d=26,10$.
Due to the low dimensionality ($d = {\hat c} + 1 \le 2$)
of the noncritical theories the remnants of the excited
states can only appear as poles in the amplitudes for certain discrete
values of the momentum. From this point
of view the situation is similar to the {\it critical} $N=2$ string
which contains only a massless scalar field (a deformation of the
Kh\"aler potential) in its spectrum in the NS-sector. Tree scattering
amplitudes of this particle have been calculated in [7] and the
expected simplicity was confirmed by the vanishing of the
four-point function. Those results suggest the study of
a possible $N=2$ {\it noncritical}
string. In particular, due to the finiteness of
the spectrum of the critical case, we do not expect
an infinite tower of discrete states in the $d\to 2^-$
limit like the $N=0,1$ noncritical strings. In a recent work\ref{8}
we made an attempt to understand noncritical $N=2$ strings
and the preliminary results that we found will be
presented and commented in the third and final section of this
talk. For sake of comparison the calculations of
[7] for the critical case are reviewed in some detail in the next section.
We finish by mentioning
some open problems and perspectives of the $N=2$ noncritical string.

\vfill\eject
\centerline {\bf 2. The Critical Case}
\vskip .7cm

\noindent The critical $N=2$ string lives in a space time with $d=2$ complex
dimensions described by the $N=2$ chiral (anti-chiral) superfields
$X^i({\overline X}^i ,i=1,2)$. On shell we have the decomposition\ref{7}
(analogously
for $\overline X^i$) :
$$
X^i=x^i(z,\overline z)+\psi_R^i(z)\theta^-+\psi _L^i(\overline z)
\overline \theta ^-
- \partial x^i \theta ^+\theta^- - \overline \partial x^i \overline \theta
^+\overline \theta^- \eqno(1)$$
where $(\theta^\pm)^\dagger =\overline \theta^\mp$. The component fields
have in our notation the following propagators:

$$
\eqalign{
\langle x^i(z)\overline x^j(w)\rangle&={\eta}^{ij}\ln \vert z-w\vert ^{-2}\cr
\langle \psi^i_R(z)\overline \psi^j_R(w)\rangle &
=\langle \psi^i_L(\oz)\overline \psi^j_L(\ow)\rangle^*
=2{\eta}^{ij}(z-w)^{-1}\cr}
\eqno(2)$$
where ${\eta}^{ij}=(+,-)$.The vertex operator below
represents the massless scalar particle
mentioned in the introduction:
$$V(k)=\int d^2zd^4\theta e^{ik\cdot \overline X(z)+i\overline
k\cdot X(z)}\eqno(3)$$
In order that $V(k)$ be a physical operator its $U(1)$ charge (q)
and conformal weight ($\Delta$) must vanish. The first requirement
is automatically satisfied by (3) and the second one imply the
on shell condition:
$$
k\cdot \overline k = k_1\overline k_1 - k_2\overline k_2 = 0.\eqno(4)$$
Now we can calculate $n$-particle amplitudes:

$$
{\cal A}_n=\langle V_{k_1}\cdots V_{k_n}\rangle =
\prod_{i=1}^n\int d^2z_id^4\theta_i
\left\langle \prod_{i=1}^N e^{i(k_i\overline X_i + \overline k_i X_i)}
\right\rangle
\eqno(5)$$
Integration over the zero-modes $x_0^i,y_0^j$ of the first component
of the supercoordinate ($x^j=x^j +iy^j$) leads to the momentum
and energy conservation:

$$\sum _{j=1}^nk_j=0=\sum _{j=1}^n{\ok}_j\eqno(6)$$
Furthermore, the residual
$OSP(2,2)$ symmetry of the superconformal gauge permits us to fix, e.g.,
$\theta_1^{(\pm)}=\theta_3^{(\pm)}=0$ and $
z_1=\infty\, ,\,, z_2=1\, ,\, z_3=0$. In this case we have for the
3-particle scattering:

$$
\eqalignno{
{\cal A}_3 = \Biggl\langle e^{i(k_3\cdot {\overline x}(0)+
{\overline k}_3\cdot x(0))}&e^{i(k_2\cdot {\overline x}(1)+
{\overline k}_2\cdot x(1))}
[ik_2\cdot \partial \overline x - i\overline k_2\cdot
\partial x - (k_2\cdot
\overline \psi_R)( \overline k_2\cdot \psi_R)]&\cr
\times &[ik_2\cdot \overline \partial \overline x - i\overline k_2\cdot
\overline \partial x - (k_2\cdot
\overline \psi_L)( \overline k_2\cdot \psi_L)]\Biggr\rangle &(7)\cr}$$
Using the propagators (2) we reproduce Ooguri and Vafa's result\ref{7}:

$${\cal A}_3 = (c_{23})^2\eqno(8)$$
where $c_{ij} = k_i\cdot \overline k_j - \overline k_i \cdot k_j$ .

For the four-particle scattering Ooguri and Vafa have
obtained\ref{7} (in the
gauge $\theta_1^{(\pm)}=\theta_4^{(\pm)}=0\, ,
z_1=\infty\, ,\, z_2=1\, ,\, z_3=z\,
,\, z_4=0$):
$$
{\cal A}_4=\!\int\!\! d^2z \vert z \vert ^{2s_{34}} \vert
1-z\vert ^{2s_{32}}\!\left\vert{s_{32}(s_{32}-1)\over (1-z)^2} \!
+ \!{c_{12}c_{34}\over z}\!+ \!
{c_{23}c_{41}\over 1-z}\right\vert^2
\eqno(9)$$
where $s_{ij}=k_i\cdot \overline k_j+\overline k_i\cdot k_j$.
The above integral can be calculated by generalizing the technique
of analytic continuation of Dotsenko\ref{9} implemented originally in
the calculation of the simpler
integral $\int\!\! d^2z \vert z \vert ^{2\alpha} \vert
1-z\vert ^{2\beta}$ (see also [10]). Finally one obtains:
$$
{\cal A}_4 = - {\pi F^2 \over 16}\Delta (1+s_{34})\Delta
(1+s_{14})\Delta (1+s_{24})
\eqno(10)$$
where $\Delta (x)=\Gamma(x)/\Gamma(1-x)$ and
$$
F= 1- {c_{23}c_{41}\over s_{14}s_{24}} -
{c_{34}c_{12}\over s_{34}s_{24}}. \eqno(11)
$$
Above, we have used the identities:
$$
s_{32}=s_{14}\quad , \quad s_{14}+s_{24}+s_{34}=0 .\eqno(12)
$$
It turns out that after use of the on shell condition (4) one gets
$F=0$ identically, and therefore:
$$
\aquatro = 0 \eqno(13)
$$
It is expected that higher point amplitudes also vanish
in the same fashion\ref{7}.

Before we finish this section note that formula (10) may be checked quite
quickly by looking at the residues of (9) even though we did not
know how to calculate exactly the complicated integrals (9).
For instance, let us calculate the residue $(R)$ of $\aquatro$ at the first
pole of the $ (34) $-channel. Supposing $s_{34}= -1+\epsilon$ we have:

$$
{R\over\epsilon}=\!\int\!\! d^2z \vert z \vert ^{-2+\epsilon} \vert
1-z\vert ^{2s_{32}}\!\left\vert{s_{32}(s_{32}-1)\over (1-z)^2} \!
+ \!{c_{12}c_{34}\over z}\!+ \!
{c_{23}c_{41}\over 1-z}\right\vert^2
\eqno(14)$$
Next we use the following representation for the distribution
$\vert z \vert^{-2+\epsilon}$:
$$
\vert z \vert^{-2+\epsilon} =
{\pi\over\epsilon}{\delta}^{(2)}(z). \eqno(15)
$$
Thus we obtain,
$$
\eqalignno{
{R\over \pi}= &(s_{32}(s_{32}-1)+c_{23}c_{41})^2 \cr
&+ (c_{34}c_{12})^2\int d^2z \partial_z\partial_{\oz}{\delta}^{(2)}(z)
\vert 1-z \vert^{2s_{32}}\cr
&+c_{34}c_{12}s_{32}(s_{32}-1)\int d^2z(-\partial_z{\delta}^{(2)}(z))
(1-z)^{s_{32}}(1-\oz)^{s_{32}-2}\cr
&+ h.c.\cr
&+c_{34}c_{12}c_{23}c_{41}\int d^2z \partial_z{\delta}^{(2)}(z)
(1-z)^{s_{32}}(1-\oz)^{s_{32}-1}&\cr
&+ h.c.\cr
&=[s_{32}(s_{32}-1)]^2\left(1+ {c_{23}c_{41}\over s_{32}(s_{32}-1)}+
 {c_{34}c_{12}\over(s_{32}-1)}\right)^2\quad . &(16)\cr}
$$
Using the identities (12) we have:
$$
R=\pi [F(s_{34}=-1)]^2 (1-s_{14})^2(s_{14})^2 = \pi [F(s_{34}=-1)]^2
\Delta(1+s_{14}) \Delta(1+s_{24})
\eqno (17)$$
This is the result expected from (10). The analysis of other poles and channels
also confirms formula (10).
\vskip .7cm
\centerline {\bf 3. The Noncritical Case }
\vskip .7cm
\noindent Similar to the $N=0,1$ noncritical $2d$-strings studied in [1,5]
it is natural to consider the coupling of a chiral (anti-chiral)
$N=2$ superfield $X(\ox)$
in a Coulomb gas representation with ${\hat c}\le 1(c_X=3{\hat c})$
to a superliouville chiral (anti-chiral) superfield $\Phi(\overline\Phi)$
such that the total action (suppressing the cosmological terms\ref{8})
is given by:

$$
\eqalign{
S =&  S_{\Phi} + S_X \cr
S_{\Phi}
=&{1\over 4\pi}\!\!\int\!\! d^2w d^4\theta\hat E\left(\Phi
\overline \Phi
-Q\hat Y(\Phi + \overline \Phi)\right)\cr
S_X =& {1\over 4\pi}\!\!\int\!\! d^2w d^4\theta \hat E\left( X\overline X +
 2i\alpha_0\hat Y(X +\overline X)\right) \cr}\eqno(18)
$$
The quantity ${\hat Y}$ stands for the $N=2$ supercurvature superfield
and $\hat E$ for the superdeterminant of the superzweibein. In order
to introduce
the notation we give the on shell decomposition
(analogous for $\overline \Phi$ , see also (1)) :
$$
\Phi=\varphi(z,\overline z)+\xi_R(z)\theta^-+\xi _L(\overline z)
\overline \theta ^-
- \partial \varphi \theta ^+\theta^- - \overline \partial
\varphi \overline \theta
^+\overline \theta^- \eqno(19)
$$
The total energy momentum tensor and $U(1)$ current are given respectively
by:

$$
\eqalign{
T&=T_{\Phi}+T_{X}\cr
T_{\Phi}&=-\colon \partial \overline \varphi\partial\varphi \colon
+{1\over 4}\colon \overline\xi_R
\partial\xi_R \colon + {1\over 4}\colon \xi_R\partial\overline\xi_R \colon
-{Q\over 2}\partial ^2(\varphi+\overline\varphi) \cr
T_X&=- \colon \partial\overline x \partial x \colon
+ {1\over 4}\colon \overline \psi_R\partial
\psi_R \colon + {1\over 4}\colon \psi_R \partial \overline\psi_R
\colon + i\alpha_0\partial ^2(x+\overline x )\cr }\eqno(20)
$$

$$\eqalign{
J&=J_{\Phi}+J_X\cr
J_{\Phi}&={1\over 4}\colon \overline\xi_R\xi_R\colon +{Q\over 2}
\partial(\varphi - \overline \varphi)\cr
J_{X}&={1\over 4}\colon \overline\psi_R\psi_R\colon
-i\alpha_0\partial( x - \overline x)\quad .
\cr }\eqno(21)$$
Analogously to (2) we have the following propagators:
$$
\eqalign{
\langle x(z)\overline x(w)\rangle&=\langle
\varphi(z)\overline \varphi(w)\rangle =\ln \vert z-w\vert ^{-2}\cr
\langle \psi_R(z)\overline \psi_R(w)\rangle &=\langle \xi_R (
z)\overline \xi_R (w) \rangle =2(z-w)^{-1}\cr
\langle \psi_L(\overline z)\overline \psi_L(\overline w)\rangle &=
\langle \xi_L (\overline z)\overline \xi_L (\overline w) \rangle
=2(\overline z-\overline w)^{-1}\cr}
\eqno(22)$$
Following [11]
we require a vanishing total central charge, fixing $Q$ to be :
$$Q=2\vert \alpha_0\vert .\eqno(23)$$
Still following [11] the noncritical version of the vertex (3)
is defined by:
$$
V(k,\overline k)=\int d^2zd^4\theta e^{i(k \overline
X+\overline k X) +\beta\overline \Phi + \overline \beta \Phi}\eqno(24)
$$
Imposing vanishing conformal weight and $U(1)$ charge we have two equations
for the
dressings $\beta$ and $\overline \beta$:
$$
\eqalignno{
\Delta\left(e^{i(k \overline X+\overline k
X)+\beta\overline\Phi+\overline\beta\Phi }\right)\!& = 2\left[(\overline
k-\alpha_0)(k-\alpha_0) -  (\overline\beta + {Q\over 2})(\beta + {Q\over 2})
\right]=0&\cr
q\left(e^{i(k \overline X+\overline k
X)+\beta\overline\Phi+\overline\beta\Phi }\right) & =
2\alpha_0 (k-\overline k)+Q(\beta - \overline\beta )=0\quad . &(25)\cr }
$$
The second equation of (25) determines the imaginary part of the dressing
 (for $\alpha_0\ne 0$) and by plugging it in the first one
we obtain the real part of the dressing (up to a sign):
$$
E_{\pm}=\left( {\beta +\overline \beta \over 2}\right) _\pm +
{Q\over 2} = \pm \left
\vert {k+\overline k\over 2} - \alpha_0\right \vert
\eqno(26)$$
where $E_{\pm}$ is the energy associated with the time direction
($\Phi + \overline\Phi$). Following Seiberg\ref{12} we take henceforth
only positive energy solutions $(E_+)$.

Having defined the vertex operator we can start calculating
$n$-point
correlation functions ${\cal A}_n =\langle V_{k_1}\cdots V_{k_n}\rangle$.
Integrating over the double zero-modes of $x$ and $\Phi$ we obtain the
momentum and energy conservation laws, respectively\newfoot{*}{We have
used that on the sphere $\int d^2z{\sqrt {\hat g}}\hat R = 8\pi$.}:
$$\sum _{j=1}^nk_j=2\alpha_0=\sum _{j=1}^n{\ok}_j\eqno(27)$$

$$
\sum _1^n\beta _i + Q =0 = \sum _1^n \overline \beta _i +  Q
\eqno(28)
$$
The calculation of the amplitudes is very similar to the critical
case and we obtain for the 3-point function:
$${\cal A}_3 = (\ln \mu )^2 (c_{23})^2\eqno(29)$$
where now,
$$
\eqalign{
c_{ij}& = k_i\cdot \overline k_j - \overline k_i \cdot k_j \cr
& = k_i \overline k_j - \beta _i\overline \beta _j
 - \overline k_i k_j+\overline \beta _i\beta _j \cr }
\eqno(30)$$
The overall factor $(\ln\mu)^2$ comes from the volume of the
Liouville zero-modes (see [8])
with $\mu$ interpreted as a cosmological constant.
Now the important difference with respect to the critical case comes
from the non analytical structure of the dispersion relation
(26) which allows us to eliminate completely, in a given kinematic region,
the real part of the momentum of one of the scattered particles and
to rewrite $\atres$ in a factorized form. For instance, in the
region\newfoot{**}{Calculations for $\alpha_0>0$ are completely analogous.}
$\Re\!e k_2\, ,\, \Re\!e k_3\le\alpha_0\, ,\,\Re\!e k_1\ge\alpha_0<0$
we have :
$$
\Re\!e k_1 =0 \quad ; \quad k_1\cdot{\ok}_1 = 0 \eqno(31)
$$
and (29) can be written as\ref{8}:
$$
{\cal A}_3 =(\ln \mu )^2 \left({\vert k_1 \vert\over
\alpha_0}\right)^2 (k_2\cdot \overline k_2)(k_3\cdot \overline k_3)
\eqno(32)$$

In the same way if we repeat now the calculation for $\aquatro$
we get at the first sight the same result of the critical case, i.e. (10),
multiplied by the factor $(\ln \mu )^2$. But in a given kinematic region,
e.g., $\Re\!e k_1$, $\Re\!e k_2$, $\Re\!e k_3 \le\alpha_0$, $\Re\!e k_4
\ge\alpha_0<0$, we have:

$$
\Re\!e k_4 = -\alpha_0 \quad ; \quad k_4\cdot{\ok}_4 = 0 \eqno(33)
$$
and after some algebra,
$$
F= {1\over4}\left({\vert k_4 \vert\over\alpha_0}\right)^2\prod_{i=1}^3
{(k_i\cdot \overline k_i)\over s_{i4}}.
\eqno(34)$$
By further noticing that, in the above kinematic region, $s_{i4}=-k_i\cdot
{\ok}_i$
we can write $\aquatro$ (see (10)) in a completely factorized form\ref{8}:
$$
{\cal A}_4 = {\pi (\ln \mu )^2\over 16}
\left({\vert k_4 \vert\over\alpha_0}\right)^4\prod _1^3
\Delta (1-k_i\cdot \overline k_i).
\eqno(35)$$
It's important to remark that in any kinematic region where at least
two particles satisfy $\Re\!e k_i\ge \alpha_0<0
\quad (\rightarrow k_i\cdot{\ok}_i=0)$
both amplitudes $\atres$ and $\aquatro$ vanish.

We can start now the analysis of the results (32) and (35) observing that
when we take $\alpha_0=0 \quad ({\hat c}=1)$ the
$U(1)$ charge of the vertex operator vanishes identically (see (25))
and we have no restrictions on the imaginary part of the dressing
($\beta - \overline\beta$). In particular, this means that the
dispersion relation (26) does not apply to the $\alpha_0=0$ case. Therefore
the factorized results that we have obtained so far are only true, strictly
speaking, for ${\hat c} < 1 \quad (\alpha_0\ne 0)$. For $\alpha_0=0$ the
requirement of vanishing conformal weight (see (25)) just reproduces
the on shell condition $k\cdot\ok=k\ok - \beta\overline\beta=0$ and we
recover the critical case whose amplitudes have been already calculated
by Ooguri and Vafa\ref{7}. The discontinuous nature of the ${\hat c}\to 1^-
(\alpha_0\to 0)$ limit can be also seen from other points of view. Note, for
instance, that contrary to the $N=0,1$ noncritical strings, in the $N=2$
case it is impossible to obtain the ${\hat c}=1$ noncritical theory
by an appropriate rotation of the ${\hat c}<1$ model (see (20),(21) and
(23)). One can also take the factorized expression
(35) for $\aquatro$ in the limit $\alpha_0\to 0$ to see that the result
diverges with ${1\over\alpha_0}$ which shows the
non existence of
discrete states in the ${\hat c}\to 1^-$ limit, as expected. For ${\hat c} < 1$
the interesting models are the minimal ones for which the functions
$\Delta (1-k\cdot\ok)$ have no poles
or zeroes. Thus, as in the $N=0,1$ noncritical strings, these functions
have a mild effect and can be absorbed through redefinitions
of the vertices.

Concluding we must say that there are still many aspects of $N=2$
noncritical strings to be understood in the continuous which might be useful
in developing super matrix models. In particular we do not
know how to continue for other kinematic regions the results that we
have derived in a given region. For this aim it may be useful
to calculate higher point functions (work in progress), as well as, to
properly include the cosmological terms\ref{8),13} to understand
the space time picture behind the amplitudes that we
have obtained.
\vskip .7cm
\centerline {\bf Acknowledgments }
\vskip .15cm
\noindent This talk is based on a work\ref{8} in collaboration with
E. Abdalla and M.C.B. Abdalla to whom I am in debt
for many useful discussions and incentive. I also acknowledge
C. Vafa for  a discussion
during the summer school and the organizers
of the Workshop on Superstrings for the opportunity to present this work.
The financial support of Fapesp is also acknowledged.
\vskip .7cm
\centerline {\bf References}
\vskip .15cm
\refer[[1]/P. Di Francesco and D. Kutasov, Nucl Phys. {\bf B375}(1992)119.]

\refer[[2]/I. Klebanov, Priceton Preprint PUPT-1271, 1991.]

\refer[[3]/M. Bershadsky and I. Klebanov, Nucl.Phys. {\bf B360}(1991)559;
S. Govindarajan, T. Jayaraman and V. John ,``Chiral Rings and Physical
States in $c<1$ String Theory'' (Madras preprint IMSc-92/30).]

\refer[[4]/M. Goulian and M. Li, Phys. Rev. Lett. {\bf 66}(1991)2051;
Vl. Dotsenko,Mod. Phys. Lett {\bf A6}(1991)3601.]

\refer[[5]/E. Abdalla, M. C. B. Abdalla, D. Dalmazi, K. Harada, Phys. Rev.
Lett. {\bf 68 }  (1992) 1641; Int. J. Mod. Phys. {\bf A } (to appear).]

\refer[[6]/L. Alvarez-Gaum\'e and J. L. Ma\~nez, Mod. Phys. Lett. {\bf A6}
(1991)2039; K. Aoki, E. D'Hoker, Mod. Phys. Lett. {\bf A7} (1992)333.]

\refer[[7]/H. Ooguri, C. Vafa, Nucl. Phys. {\bf B361} (1991)469, {\bf B367}
(1991)83.]

\refer[[8]/E. Abdalla, M.C.B. Abdalla and D. Dalmazi, Phys. Lett {\bf B}
(to appear), see also S\~ao Paulo Univ. preprint IFUSP/P-998.]

\refer[[9]/Vl. Dotsenko, Advanced Studies in Pure Mathematics {\bf 16}(1988)
123.]

\refer[[10]/E. D'Hoker and D.H. Phong, Rev. of Mod. Phys. {\bf 60}(1988)917;
H. Kawai, D.C. Lewellen and S.H.-Tye, Nucl. Phys. {\bf B269} (1986)1.]

\refer[[11]/J. Distler, Z. Hlouzek and H. Kawai, Int. J. of Mod. Phys.
{\bf A5}(1990)391.]

\refer[[12]/N. Seiberg, Lecture at 1990 Yukawa Int. Sem. Common Trends in Math.
and Quantum Field Theory, and Cargese meeting Random Surfaces, Quantum Gravity
and Strings, May 27, June 2, 1990.]

\refer[[13]/N. Seiberg and D. Kutasov, Phys. Lett. {\bf B251}(1990)67.]

\end